# (CaO)(FeSe): A Layered Wide Gap Oxychalcogenide Semiconductor

Fei Han,[†] Di Wang,[‡] Christos D. Malliakas,[†,§] Mihai Sturza,[†] Duck Young Chung,[†] Xiangang Wan,[‡] and Mercouri G. Kanatzidis[*,†,§]

[†] Materials Science Division, Argonne National Laboratory, Argonne, Illinois 60439, United States
[‡] National Laboratory of Solid State Microstructures, School of Physics, Collaborative Innovation Center of Advanced Microstructures, Nanjing University, Nanjing 210093, China
[§] Department of Chemistry, Northwestern University, Evanston, Illinois 60208, United States

**ABSTRACT:** A new iron-oxychalcogenide (CaO)(FeSe) was obtained which crystallizes in the orthorhombic space group *Pnma* (No. 62) with $a$ = 5.9180(12) Å, $b$ = 3.8802(8) Å, $c$ = 13.193(3) Å. The unique structure of (CaO)(FeSe) is built up of a quasi-two-dimensional network of corrugated infinite layers of corner-shared FeSe$_2$O$_2$ tetrahedra that extend in the *ab*-plane. The corrugated layers composed of corner-shared FeSe$_2$O$_2$ tetrahedra stack along the *c*-axis with Ca$^{2+}$ cations sandwiched between the layers. Optical spectroscopy and resistivity measurements reveal semiconducting behavior with an indirect optical band gap of around 1.8 eV and an activation energy of 0.19(1) eV. Electronic band structure calculations at the density function level predict a magnetic configuration as ground state and confirm the presence of an indirect wide gap in (CaO)(FeSe).

## INTRODUCTION

The mixed anion systems such as oxypnictides and oxychalcogenides have shown fascinating properties highlighted by the discovery of iron-based superconductors with $T_c$ up to 56 K.[1-5] These systems usually adopt layered structures allowing segregation of the two anions, owing to their different chemical nature.[6] In the iron-oxypnictide superconductors $Ln$FeAsO$_{1-x}$F$_x$ ($Ln$ = rare earth),[1-5] Sr$_2$ScO$_3$FeP,[7] and Sr$_2$VO$_3$FeAs,[8] Fe$X$ ($X$ = pnictide) layers of antifluorite-like edge-shared Fe$X_4$ tetrahedra are responsible for the superconductivity while oxide spacer layers act as charge reservoirs which can be doped. Since β-FeSe (under high pressure),[9-11] $K_x$Fe$_{2-y}$Se$_2$,[12-14] and FeSe with molecular interlayers[15-17] were observed to be superconducting above 30 K, the exploration of new iron-oxychalcogenide compounds has intensified. In the past several years, a series of new iron-oxychalcogenides including Na$_2$Fe$_2$Se$_2$O,[18] $Ae$Fe$_2$Q$_2$O ($Ae$ = Sr, Ba, Q = S, Se),[19-22] β-La$_2$O$_2$$M$Se$_2$ ($M$ = Mn, Fe),[23] Ce$_2$O$_2$FeSe$_2$,[24] and (BaF)$_2$Fe$_{2-x}$Q$_3$ ($Q$ = S, Se)[25] have been reported. These materials along with the previously reported $Ln_2$O$_2$Fe$_2$OSe$_2$[26-29] and $Ae_2$F$_2$Fe$_2$OSe$_2$[30] ($Ae$ = alkaline earth) are not superconductors. Coincidentally they do not contain isostructural FeSe layers of antifluorite-like edge-shared FeSe$_4$ tetrahedra. In contrast the recently reported (Li$_{0.8}$Fe$_{0.2}$)OHFe$_2$Se$_2$ has a coexistence of superconductivity at 42 K and antiferromagnetism at 8.6 K,[31,32] and it consists of FeSe conductive layers providing superconductivity and (Li$_{0.8}$Fe$_{0.2}$)OH spacer layers in which at the shared Li/Fe site the observed antiferromagnetic ordering possibly happens.

In an effort to synthesize a new analogue containing FeSe layers and CaO spacer layers for a new iron-based superconductor which was also aimed by other groups,[33] we discovered the new compound (CaO)(FeSe) which crystallizes in the orthorhombic space group *Pnma* (No. 62) with $a$ = 5.9180(12) Å, $b$ = 3.8802(8) Å, $c$ = 13.193(3) Å. (CaO)(FeSe) adopts a new type of layered structure that remarkably features O$^{2-}$ and Se$^{2-}$ anions in the same layer. (CaO)(FeSe) consists of corrugated slabs of corner-shared FeSe$_2$O$_2$ tetrahedra that extend along the *ab*-plane and stack along the *c*-axis with Ca$^{2+}$ cations sandwiched between the FeSe$_2$O$_2$ slabs. This new material is a wide band gap semiconductor with an indirect optical gap of around 1.8 eV.

## EXPERIMENTAL SECTION

**Sample Preparation.** Single-crystalline (CaO)(FeSe) was grown from the melt. Our initial target was to prepare CaFe$_2$Se$_2$O with the following procedure: Precursor FeSe was firstly synthesized by direct combination of powdered Fe (Alfa Aesar, 99.9+% metals basis) and Se (Alfa Aesar, 99.999% metals basis). Starting materials of CaO powders (Sigma-Aldrich, 99.9% trace metals basis) and the pre-reacted FeSe with the molar ratio of 1:2 and total mass of around 3 g were mixed and placed in an alumina crucible. All handling was performed in a glove box under Argon atmosphere (both H$_2$O and O$_2$ are limited below 0.1 ppm). The alumina crucible was jacketed by an evacuated silica tube. The tube was heated to 1000 °C in a box furnace and kept at 1000 °C for 10 hours. A slow-cooling process from 1000 °C to 700 °C was carried out during 60 hours. From the powder X-ray diffraction data, we found the resultant product consisted of CaSe, FeSe, FeO, and an unindexed phase. With the help of an optical microscope, we found some red transparent crystals covered by black melt in the product. A hand-picked crystal isolated from the black matrix is shown Figure 1(a). These red transparent crystals were finally determined as (CaO)(FeSe) by single-crystal X-ray diffraction. After the stoichiometry of the new quaternary compound was elucidated, we tried a more rational synthesis of (CaO)(FeSe) by direct combination of a stoichiometric mixture of CaO and FeSe. The heating procedure was the same as above. However, the product was impure and consisted of CaSe, FeSe, $Ca_2Fe_2O_5$, $CaFe_3O_5$, and (CaO)(FeSe). The impure nature of the samples and the low yield of (CaO)(FeSe) phase made bulk measurements such as magnetic susceptibility and heat capacity impossible.

**Scanning Electron Microscopy.** Elemental analysis of (CaO)(FeSe) crystal was performed on a Hitachi S-4700-II Scanning Electron Microscope with Energy Dispersive Spectroscope (EDS) detector equipped. Electron micrograph in the inset of Figure 1(b) shows the plate-like crystal habit of a single (CaO)(FeSe) crystal. Typical crystal dimensions were around 150 × 150 × 5 μm³. Semiquantitative analysis by EDS indicated the crystal consisted of Ca, Fe, Se, and O, as shown in Figure 1(b), with an average composition of $Ca_{0.22(1)}Fe_{0.23(1)}Se_{0.24(1)}O_{0.30(2)}$.

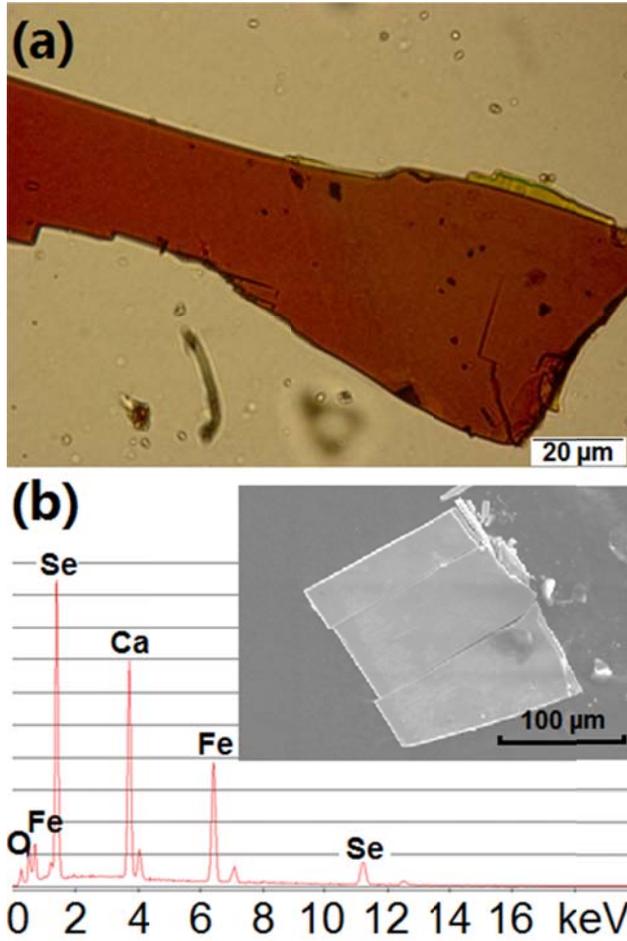

**Figure 1.** (a) Photograph of a red (CaO)(FeSe) transparent crystal. (b) Energy Dispersive Spectrum (EDS) collected on a typical (CaO)(FeSe) crystal whose electron micrograph is shown in the inset.

**Single-crystal X-ray Diffraction.** Single-crystal X-ray diffraction measurement at room temperature was carried out on a STOE diffractometer. Data reducing, integration and absorption correction were performed with the software X-Area,[34] and the structure was solved by direct methods and refined using the SHELXTL software.[35]

**Optical Spectroscopy.** Single-crystal absorption spectrum was obtained on a Hitachi U-6000 Microscopic FT spectrophotometer mounted on an Olympus BH2-UMA microscope. Crystal lying on a glass slide was positioned over the light source and the transmitted light was detected from above. The background signal of the glass slide was subtracted from the collected intensity.

**Table 1. Single Crystal Data and Structure Refinement for (CaO)(FeSe) Collected at Room Temperature.**

| Empirical formula | (CaO)(FeSe) |
|---|---|
| Formula weight | 190.89 |
| Wavelength | 0.71073 Å |
| Crystal system | Orthorhombic |
| Space group | Pnma |
| Unit cell dimensions | a = 5.9180(12) Å, α = 90.00°<br>b = 3.8802(8) Å, β = 90.00°<br>c = 13.193(3) Å, γ = 90.00° |
| Volume | 302.95(11) Å³ |
| Z | 4 |
| Density (calculated) | 4.185 g/cm³ |
| Absorption coefficient | 18.383 mm⁻¹ |
| F(000) | 352 |
| Crystal size | 0.15 × 0.1 × 0.005 mm³ |
| θ range for data collection | 3.77 to 25.46° |
| Index ranges | -7≤h≤7, -4≤k≤4, -15≤l≤15 |
| Reflections collected | 1563 |
| Independent reflections | 315 [$R_{int}$ = 0.0303] |
| Completeness to θ = 24.98° | 96% |
| Refinement method | Full-matrix least-squares on $F^2$ |
| Data / restraints / parameters | 315 / 0 / 25 |
| Goodness-of-fit | 1.143 |
| Final R indices [>2σ(I)] | $R_{obs}$ = 0.0447, $wR_{obs}$ = 0.1096 |
| R indices [all data] | $R_{all}$ = 0.0531, $wR_{all}$ = 0.1133 |
| Largest diff. peak and hole | 3.338 and -1.140 e·Å⁻³ |

$R=\Sigma||F_o|-|F_c||/\Sigma|F_o|$, $wR=\{\Sigma[w(|F_o|^2-|F_c|^2)^2]/\Sigma[w(|F_o|^4)]\}^{1/2}$ and $w=1/[\sigma^2(F_o^2)+(0.0770P)^2]$ where $P=(F_o^2+2F_c^2)/3$

**Table 2. Atomic Coordinates (×10⁴) and Equivalent Isotropic Displacement Parameters (Å²×10³) for (CaO)(FeSe) at Room Temperature.**

| Atom | x | y | z | $U_{eq}$* |
|---|---|---|---|---|
| Se | 2430(2) | 2500 | 9040(1) | 9(1) |
| Fe | 1125(2) | 2500 | 2929(2) | 13(1) |
| Ca | 2484(3) | 2500 | 5806(2) | 9(1) |
| O | 2919(13) | 2500 | 1745(6) | 11(2) |

*$U_{eq}$ is defined as one third of the trace of the orthogonalized $U_{ij}$ tensor

**Resistivity.** Temperature-dependent resistivity measurement was done on a Quantum Design physical property measurement system (PPMS). Considering the small dimensions of the crystal and the wide optical gap, we performed the resistivity measurement in two probe geometry in an arbitrary direction in the ab-plane using 14 μm gold wire and silver paste for making the electrical contacts.

**Band Structure Calculations.** The electronic band structure calculations have been carried out by using the full potential linearized augmented plane wave method as implemented in WIEN2K package.[36] Local spin density approximation (LSDA) for the exchange-correlation potential[37] has been used here. A 9×13×4 k-point mesh is used for the Brillouin zone integral. The self-consistent calculations are considered to be converged when the difference in the total energy of the crystal does not exceed 0.1 mRy and that in the total electronic charge does not exceed $10^{-3}$ electronic charge at consecutive steps.

## RESULTS AND DISSCUSSION

**Crystal Structure**. The layered structure of (CaO)(FeSe) projected onto the *ac*-plane is depicted in Figure 2(a). The structure is built up of a quasi-two-dimensional network of corner-shared $FeSe_2O_2$ tetrahedra along *ab*-plane, as shown in Figure 2(b). Ca ions are sandwiched between the $FeSe_2O_2$ infinite layers which stack along the *c*-axis to form the (CaO)(FeSe) structure. The iron oxychalcogenide tetrahedra are perfectly ordered with $O^{2-}$ and $Se^{2-}$ occupying their own individual sites (Wyckoff position 4c). The connectivity of the corner-shared $FeSe_2O_2$ tetrahedra through the O/Se anions is directional with levorotatory and dextrorotatory Fe-O-Fe bridges alternatively replicating along the *a*-axis and with Fe-Se-Fe bridges infinitely replicating along the *b*-axis. $Ca^{2+}$ cations connect with $O^{2-}$ anions and form zig-zag $[Ca-O]_\infty$ chains along the *b*-axis. The unique feature of the (CaO)(FeSe) structure is that unlike most of other mixed anion systems which have the two types of anions segregated in different layers, (CaO)(FeSe) combines $O^{2-}$ and $Se^{2-}$ anions in the same layer. This structure is somewhat like that of $AeFe_2Q_2O$ (Ae = Sr, Ba, Q = S, Se)[19-22] which also contain the structure fragment of Fe-O-Fe bridge levorotatory-and-dextrorotatory-alternatively flipping. In $AeFe_2Q_2O$ (Ae = Sr, Ba, Q = S, Se)[19-22] the Fe-O-Fe bridges are glued by $Se^{2-}$ anions. Differently, in (CaO)(FeSe) the Fe-O-Fe bridges connect with each other by sharing $Fe^{2+}$ cations. The crystallographic data, selected bond distances, and bond angles for (CaO)(FeSe) are listed in Tables 1-4.

**Table 3. Anisotropic Displacement Parameters ($Å^2 \times 10^3$) for (CaO)(FeSe) at Room Temperature.**

| Atom | $U_{11}$ | $U_{22}$ | $U_{33}$ | $U_{12}$ | $U_{13}$ | $U_{23}$ |
|---|---|---|---|---|---|---|
| Se | 7(1) | 8(1) | 12(1) | 0 | 1(1) | 0 |
| Fe | 12(1) | 15(1) | 13(1) | 0 | 4(1) | 0 |
| Ca | 9(2) | 7(2) | 11(2) | 0 | 1(1) | 0 |
| O | 12(3) | 11(4) | 11(4) | 0 | 1(3) | 0 |

The anisotropic displacement factor exponent takes the form: $-2\pi^2[h^2a^{*2}U_{11} + ... + 2hka^*b^*U_{12}]$.

**Table 4. Bond Lengths [Å] and Bond Angles [°] for (CaO)(FeSe) at Room Temperature.**

| Atom-Atom | Bond Length (Å) |
|---|---|
| Fe-O | 1.888(8) |
| Fe-O | 1.945(8) |
| Fe-Se(×2) | 2.5776(14) |
| Ca-O(×2) | 2.314(5) |
| Ca-Se | 2.934(2) |
| Ca-Se | 2.998(2) |
| Ca-Se(×2) | 3.032(2) |
| Atom-Atom-Atom | Bond Angle (°) |
| O-Fe-O | 137.0(3) |
| O-Fe-Se(×2) | 106.49(15) |
| O-Fe-Se(×2) | 101.40(17) |
| Se-Fe-Se | 97.65(7) |
| O-Ca-O | 113.9(4) |
| O-Ca-Se(×2) | 93.8(2) |
| O-Ca-Se(×2) | 82.00(19) |
| O-Ca-Se(×2) | 82.92(18) |
| O-Ca-Se(×2) | 161.7(2) |
| Se-Ca-Se(×2) | 92.10(6) |
| Se-Ca-Se(×2) | 93.96(6) |
| Se-Ca-Se | 79.56(7) |
| Se-Ca-Se | 172.12(11) |

There is only one crystallographically independent site for each atom in the asymmetric unit cell. Each $Fe^{2+}$ cation is tetrahedrally coordinated by two $O^{2-}$ and two $Se^{2-}$ anions while each $Ca^{2+}$ cation is octahedrally coordinated by two $O^{2-}$ and four $Se^{2-}$ anions, as shown in Figure 3(a) and 3(b). All interatomic bond lengths are within normal range, Table 4. Along the *a*-axis, $Fe^{2+}$ cations and $O^{2-}$ anions form a helical structure with alternating short (1.888(8) Å) and long (1.945(8) Å) bond lengths, Figure 3(c). In the helical $[Fe-O]_\infty$ chains the distance of two most neighbored Fe atoms is 3.1678(19) Å while that distance between the neighbor $[Fe-O]_\infty$ chains is 3.8802(8) Å which are relatively short and expected to give strong Fe–Fe interactions. In contrast, the distance of two most neighbored Fe atoms between the neighbor $FeSe_2O_2$ layers is 5.9504(25) Å which could lead to a weak magnetic coupling between the layers. The $FeSe_2O_2$ tetrahedra and $CaSe_4O_2$ octahedra are distorted because of the different chemical nature of $Se^{2-}$ and $O^{2-}$ anions. In the $FeSe_2O_2$ tetrahedra and the $CaSe_4O_2$ octahedra, anion-cation-anion bond angles are greatly different from each other. For example, in the $FeSe_2O_2$ tetrahedra O-Fe-O bond angle of 137.0(3)° is much larger than Se-Fe-Se one of 97.65(7)° and both of them are far from the ideal angle of 109.5° for a regular tetrahedron.

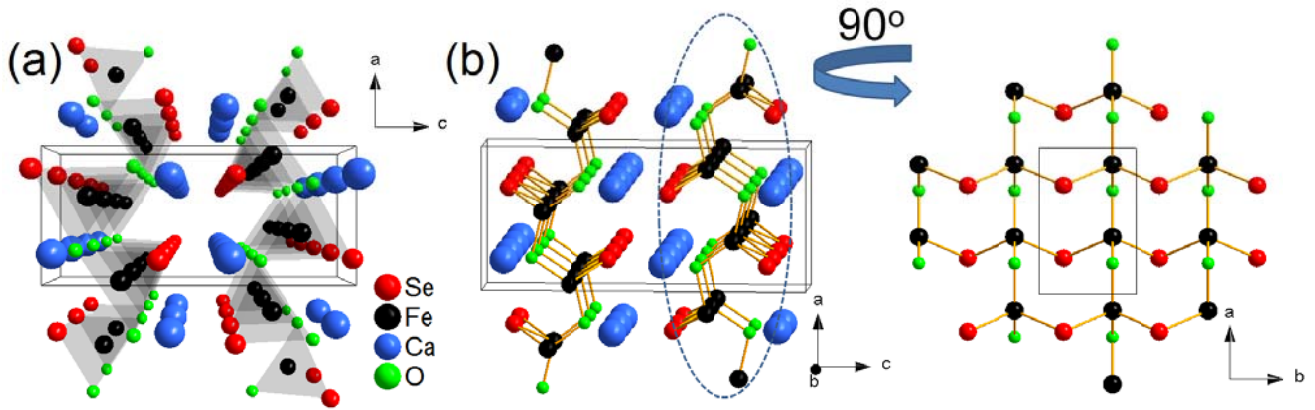

**Figure 2.** (a) Perspective view of the layered structure of (CaO)(FeSe) along crystallographic *b*-axis. (b) Bond/stick model for (CaO)(FeSe).

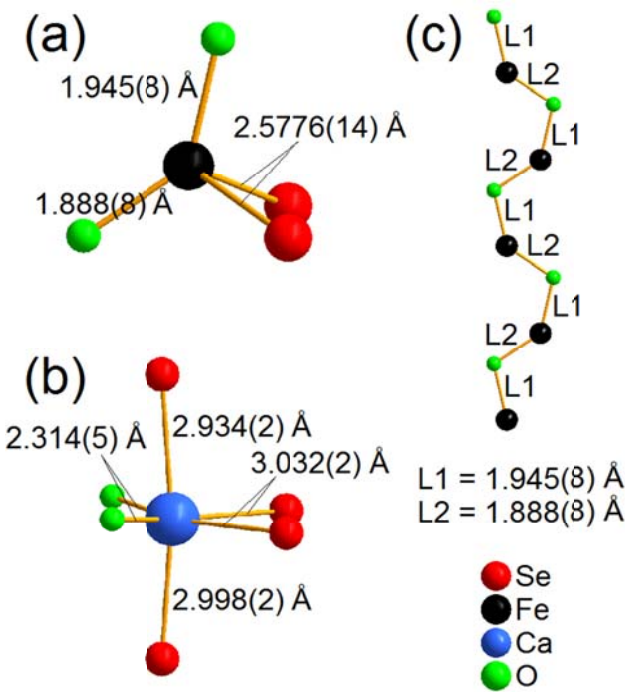

**Figure 3.** (a) Coordination environment of Fe. (b) Coordination environment of Ca. (c) [Fe-O]$_\infty$ infinite chain viewed along the *a*-axis, Fe$^{2+}$ cations and O$^{2-}$ anions form a helical structure with alternating short and long Fe-O bond lengths.

**Optical Absorption.** Based on the charge balanced formula of Ca$^{2+}$Fe$^{2+}$Se$^{2-}$O$^{2-}$ the compound is valence precise and is expected to be a semiconductor. Single-crystal absorption spectrum collected for (CaO)(FeSe) at room temperature reveal a broad band gap transition between 1.8 eV and 2.3 eV, Figure 4. For a direct optical transition the square of absorbance ($\alpha$) is expected to vary linearly with energy $\hbar\omega$ as $\alpha=(\hbar\omega-E_D)^{1/2}$ where $E_D$ is the direct optical band gap. For an indirect transition the square root of $\alpha$ is expected to vary linearly with energy as $\alpha=(\hbar\omega-E_{ID})^2$ where $E_{ID}$ is the indirect band gap energy. Since the band gap of (CaO)(FeSe) is predicted in the following subsection to be indirect we plotted the square root of absorbance data as a function of energy (inset of Figure 4) and extracted an indirect band gap value of 1.83(2) eV. This wide energy gap is consistent with the red color of the crystals.

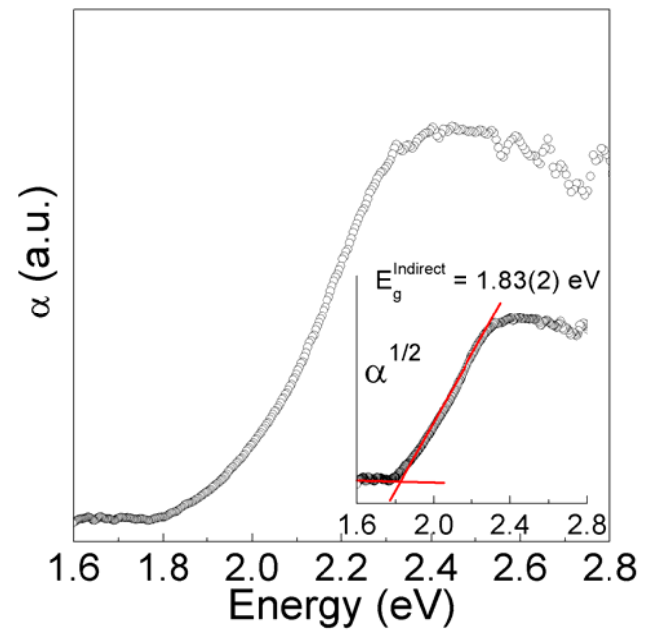

Figure 4. Absorption spectrum on a single crystal of (CaO)(FeSe) measured at room temperature, showing an indirect band gap of 1.83(2) eV (inset).

**Resistivity.** Temperature dependence of resistivity for a single crystal sample of (CaO)(FeSe) confirmed the semiconducting nature of the compound. The resistivity is around 15kΩ.cm at room temperature and increases gradually with falling temperature, Figure 5(a). Furthermore, above 256 K the temperature dependence of resistivity strictly obeys thermally activated behavior as $\rho = \rho_o\exp(E_a/k_BT)$ where $\rho_o$ refers to a prefactor and $k_B$ is the Boltzmann's constant, as shown in Figure 5(b). Within this fitting process, the activation energy $E_a$ was extracted to be 0.19(1) eV. The activation energy is much smaller than the experimental band gap indicating the presence of mid-gap levels (likely to be the partially filled $d$ levels of high spin $Fe^{2+}$) which can thermally donate electrons to conduction band or accept electrons from the valence band.

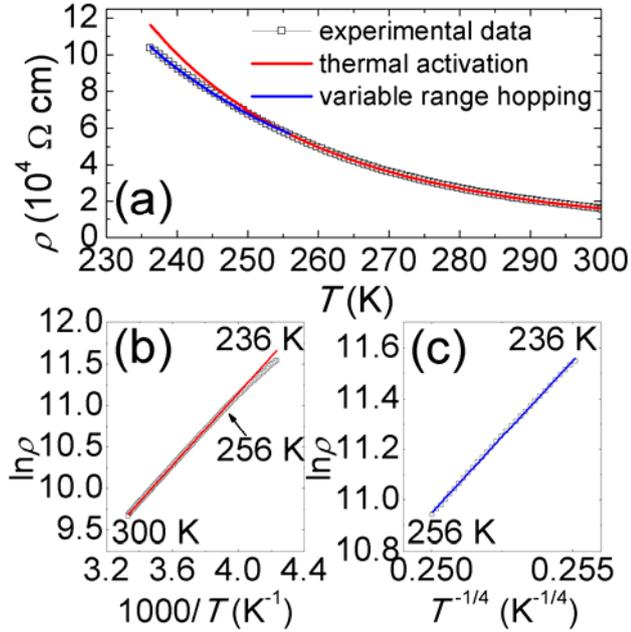

**Figure 5.** (a) Resistivity as a function of temperature for a (CaO)(FeSe) single crystal. The experimental data obey thermal activation behavior at high temperatures and 3D variable range hopping behavior at low temperatures, respectively. (b) Arrhenius plot $\ln\rho$ vs $1000/T$ showing linear behavior between 256 K and 300 K. (c) Arrhenius plot $\ln\rho$ vs vs $T^{-1/4}$ showing linearity with $T$ less than 256 K.

When $1000/T$ is more than 3.9 $K^{-1}$ ($T < 256$ K), the $\ln\rho$ vs $1/T$ curve begins to deviate from the original linearity. Initially we attempted to attribute this deviation to a smaller activation energy $E_a$ at lower temperatures, however, below 256 K the $\ln\rho$ vs $1000/T$ curve does not obey strict linearity. Instead, three-dimensional variable range hopping (3D-VRH) can better describe the resistivity behavior at low temperatures. We can see a new linear behavior in $\ln\rho$ vs $T^{-1/4}$ curve, Figure 5(c), which indicates below 256 K the 3D-VRH behavior as $\rho = \rho_o\exp(T_o/T)^{1/4}$ is obeyed, here $\rho_o$ and $T_o$ refer to prefactors. The different resistivity behaviors above 256 K and below 256 K reveal that hopping conduction dominates at low temperatures and band conduction at high temperatures.

The existence of 3D-VRH behavior in other layered transition metal compounds has been reported, which somehow indicates the strong correlation between layers.[38-40] The variation of the dominant conduction mechanism with temperature can be explained with a model of two kinds of parallel conductive channels. When the temperature is high enough, electrons are activated across the barriers between the mid-gap energy levels and conduction bands or between valence bands and the mid-gap energy levels. As the temperature falls, the thermally activated electrons decay rapidly, and at low temperatures electrons or holes choose to move through localized $Fe^{2+}$ sites with longer distance but closer energy by a tunneling process stimulated by phonons.

**Band Structure Calculations.** In the structure of (CaO)(FeSe), the Fe sublattice can be considered two staggered distorted square nets, as shown in Figure 6(a). Fe1, Fe2, Fe3, Fe4 are labeled for the representation analysis of magnetic configurations. To search for the magnetic ground state configuration, we considered five AFM configurations besides the FM states as follows (Figure 6(b)): AFM-1 (Fe1 and Fe2, Fe3 and Fe4 couple ferromagnetically while Fe1 and Fe3 have opposite spin orientation, and neighbor [Fe-O]$_\infty$ chains in one layer couple ferromagnetically), AFM-2 (Fe1 and Fe3, Fe2 and Fe4 couple ferromagnetically while Fe1 and Fe2 have opposite spin orientation, and neighbor [Fe-O]$_\infty$ chains in one layer couple ferromagnetically), AFM-3 (Fe1 and Fe4, Fe2 and Fe3 couple ferromagnetically while Fe1 and Fe2 have opposite spin orientation, and neighbor [Fe-O]$_\infty$ chains in one layer couple ferromagnetically), AFM-4 (Fe1, Fe2, Fe3 and Fe4 couple ferromagnetically while neighbor [Fe-O]$_\infty$ chains in one layer couple antiferromagnetically) and AFM-5 (Fe1 and Fe3, Fe2 and Fe4 couple ferromagnetically while Fe1 and Fe2 have opposite spin orientation, and neighbor [Fe-O]$_\infty$ chains in one layer couple antiferromagnetically).

The LSDA calculation predicts that the AFM-2 state is the ground state, and the simulated 3.0 $\mu_B$ magnetic moment mainly locates at the Fe site. As we can see from Table 5, the total energy of the AFM-3 state is only 20 meV higher than the ground state which indicates that the magnetic coupling between the two layers is small, while the total energy of other configurations are about 0.2 to 0.7 eV higher than the ground state.

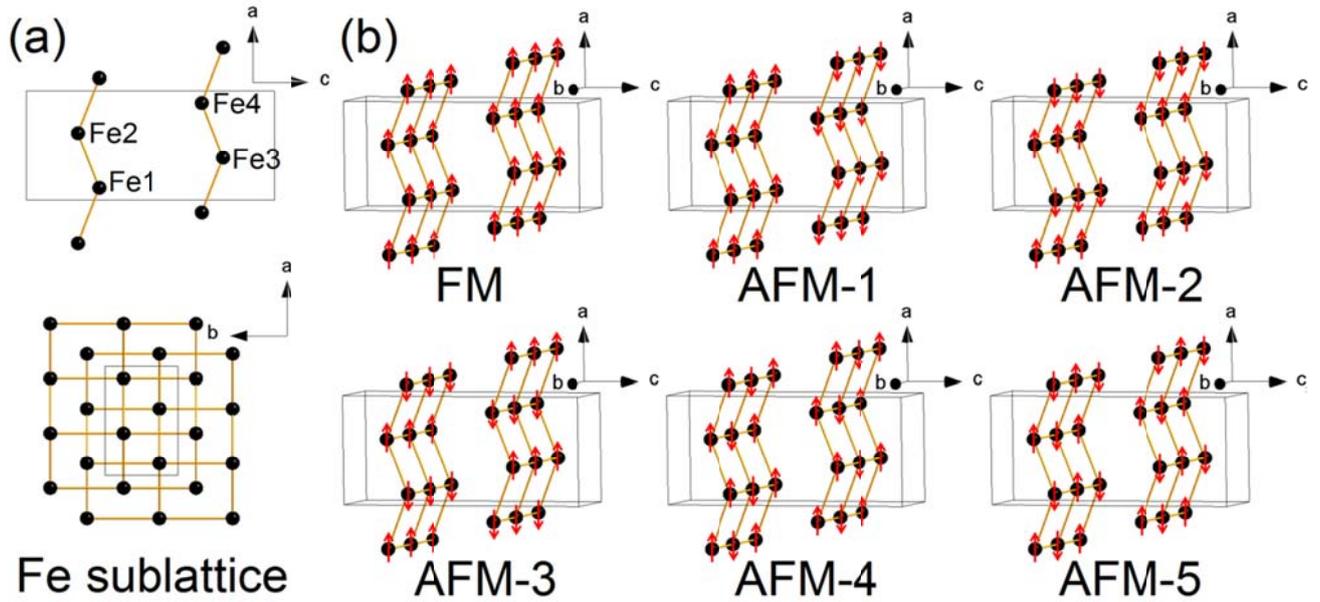

**Figure 6.** (a) Fe sublattice in the structure of (CaO)(FeSe). Fe1, Fe2, Fe3, Fe4 are labeled for the representation analysis of magnetic configurations. (b) Schematic representations of eight spin configurations including the ferromagnetic configuration FM and five different antiferromagnetic configurations AFM-1, AFM-2, AFM-3, AFM-4, AFM-5 which are described above.

**Table 5. The Total Energy [meV] of Six Magnetic Configurations Calculated by Different Scheme, Compared with the Ground State.**

|        | FM  | AFM-1 | AFM-2 | AFM-3 | AFM-4 | AFM-5 |
|--------|-----|-------|-------|-------|-------|-------|
| LSDA   | 522 | 516   | 0     | 20    | 742   | 244   |
| LSDA+$U$ | 355 | 341 | 0     | 5     | 364   | 24    |

It is well known that the electronic correlations are important in 3$d$ orbitals, so we utilized an LSDA+$U$ ($U$ = 4 eV) scheme, which is adequate for the magnetically ordered insulating ground states. As with the LSDA calculations, the LSDA+$U$ also predicts AFM-2 as the ground state, with an obtained magnetic moment of ~3.3 $\mu_B$ similar to the LSDA calculations.

The band structures of (CaO)(FeSe) from the LSDA+$U$ calculations are presented in Figure 7 and the density of states in Figure 8. The band structures show (CaO)(FeSe) is an insulator with an indirect gap of about 1.55 eV, consistent with the experimental results of 1.8 eV. The energy range, -3.0 to 3.0 eV is dominated by Fe 3$d$ states. The 12 bands located from -6.0 to -4.0 eV basically come from O 2$p$ sates, while the 12 Se 4$p$ states distribute from -4.0 to -1.0 eV. All these indicate strong hybridization between Fe 3$d$ states and Se 4$p$ sates. The Ca 4$s$ states are located about 3.0 to 5.0 eV higher than the Fermi level.

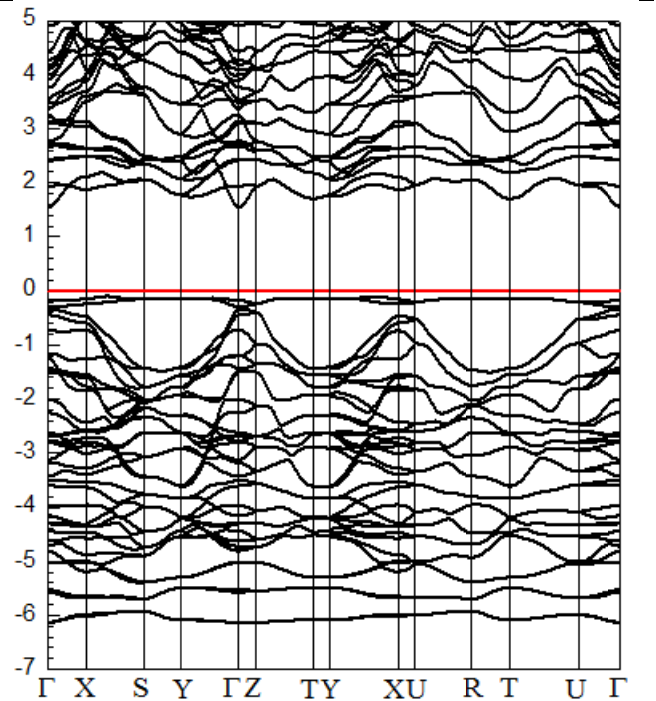

**Figure 7.** Band structures of (CaO)(FeSe) calculated by the method of LSDA+$U$ ($U$ = 4 eV). The Fermi energy is set to zero.

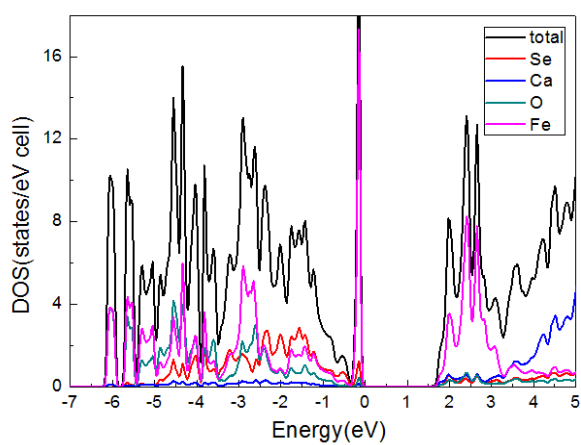

**Figure 8.** Density of states (DOS) of (CaO)(FeSe) calculated by the method of LSDA+$U$ ($U$ = 4 eV). The Fermi energy is set to zero.

## CONCLUDING REMARKS

We have successfully prepared a new iron-oxychalcogenide (CaO)(FeSe) which adopts a new structure type. The structure of (CaO)(FeSe) is unique with a combination of $O^{2-}$ and $Se^{2-}$ anions in the same $FeSe_2O_2$ layers. The structure is different from that of CaOFeS[41,42] (CaOFeS adopts a CaOZnS-type hexagonal structure) and our phase diagram studies of BaO-FeSe and SrO-FeSe only yielded $BaFe_2Se_2O$[19,20] and $SrFe_2Se_2O$[22] respectively. Therefore, the structure of (CaO)(FeSe) is stable for a restricted range of chalcogen and alkali-earth-metal elements.

## ASSOCIATED CONTENT

**Supporting Information**

X-ray crystallographic data (CIF). This material is available free of charge via the Internet at http://pubs.acs.org.

## AUTHOR INFORMATION

**Corresponding Author**

*E-mail: m-kanatzidis@northwestern.edu

**Notes**

The authors declare no competing financial interest.

## ACKNOWLEDGMENT

This work was supported by the U.S. Department of Energy, Office of Science, Basic Energy Sciences, Materials Sciences and Engineering Division. Use of the Center for Nanoscale Materials, including resources in the Electron Microscopy Center, was supported by the U. S. Department of Energy, Office of Science, Office of Basic Energy Sciences, under Contract No. DE-AC02-06CH11357. Work done at Nanjing University (by D. W. and X. W.) was supported by the NSF of China (Grants No. 11374137, 91122035 and 11174124).

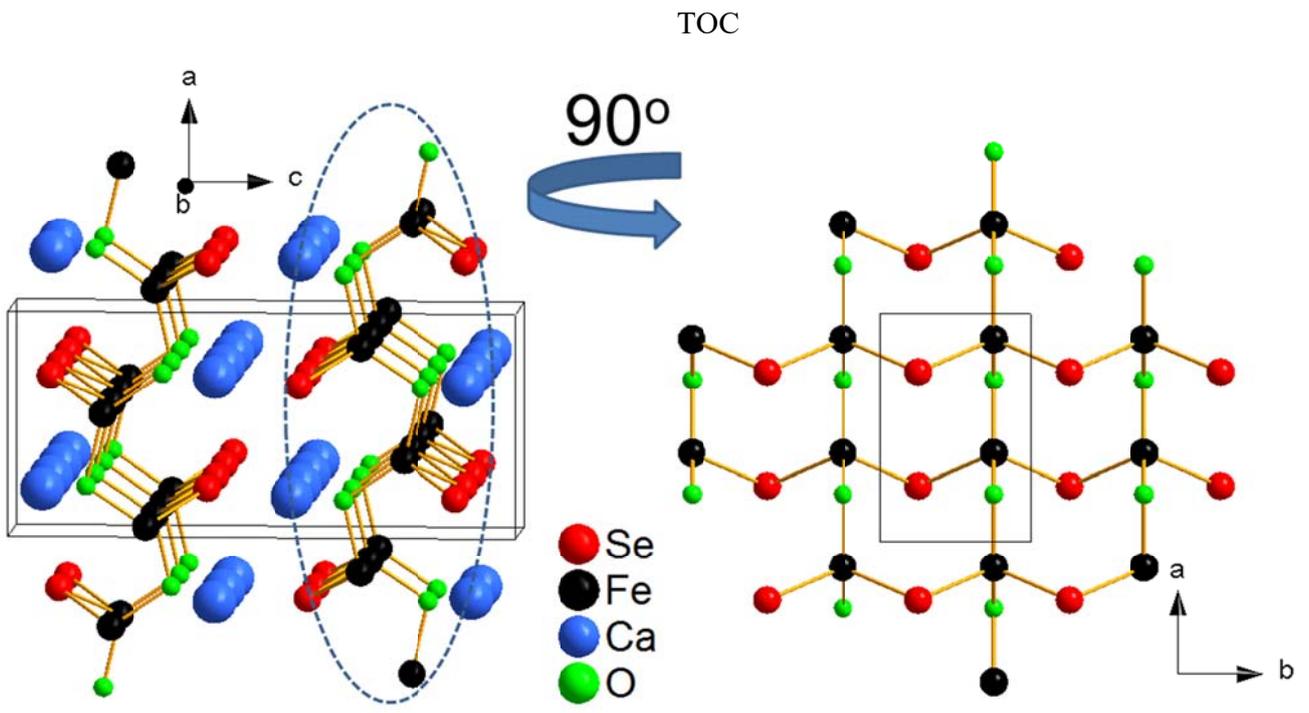